\pdfoutput=1
\documentclass[a4paper,12pt]{article}
\usepackage[utf8]{inputenc}

\usepackage[T1]{fontenc}
\usepackage{mathptmx}

\usepackage{fullpage, graphicx}
\graphicspath{{./Images/}}
\usepackage{siunitx}
\usepackage[backend=biber,sorting=none,style=nature]{biblatex}
\usepackage{url}
\usepackage{float}
\usepackage{multicol}
\usepackage{setspace}
\usepackage[symbol]{footmisc}
\usepackage{indentfirst}
\usepackage{chemformula}
\usepackage{hyperref}
\usepackage{subfigure}
\usepackage{caption}
\usepackage{mathtools}
\usepackage{setspace}
\usepackage[symbol]{footmisc}

\hypersetup{
    colorlinks=true,
    linkcolor=black,
    filecolor=black,      
    urlcolor=black,
    citecolor=black
}
\DeclareSIUnit\atm{atm}
\begin{document}

\pagestyle{empty}                       % No numbers of title page                                              % Size of crest

%\par\noindent                                           % Centre Title, and name

\begin{center}
        \LARGE\bf 
        \uppercase{Applications of 10 MeV Photons and 100 MeV Protons in Radiotherapy}\\ [20pt]      % Change to suit
\end{center}
%\vspace*{0.01cm}
\begin{center}
        \renewcommand{\thefootnote}{\fnsymbol{footnote}}
        \large Austin A. Morris
        \footnote[1]{Supervised by Paul Clegg and Gary Smith.}
        \\                    % Replace with your name
        \vspace{2mm}
        \small\it School of Physics and Astronomy, University of Edinburgh, Contact: morrisaustin11@gmail.com                                    % Submission Date
\end{center}
\vspace*{0.75mm}

%\begin{abstract}
\begin{spacing}{0.8}
\renewcommand{\baselinestretch}{0.75}
\noindent {\footnotesize\textbf{Abstract}–-Geant4 software is used to study 10 MeV photon beams and 100 MeV proton beams. Conclusions are drawn as to the advantages and disadvantages of each therapy, with some additional commentary on the secondary effects of radiation damage in treatment.}
\end{spacing}
%\end{abstract}

                                  % Change to suit
%
%                       End of Title Page
%\tableofcontents                                % Makes Table of Contents

\pagestyle{plain}                               % Page numbers at bottom
\setcounter{page}{1}   % Set page number to 1

\vspace{-2mm}
\section{Introduction}

Radiotherapy is becoming an increasingly popular method for treating cancer patients with malignant tumor growth. Two non-exclusive types of beam therapy––\textit{photon} and \textit{proton}––are used for different clinical applications at varying costs. The theoretical background of both therapies will be presented and consequently tested by a computer simulation measuring intensity and distribution of dose.

\vspace{-3mm}
\subsection{Clinical Radiotherapy}

The goal of any targeted radiotherapy is to do irreparable damage to primary cancer cells. This is achieved by double-strand breaks to malignant cell DNA, sparing healthy tissue as much as possible. The best treatments are generally achieved by \textit{fractionation}, which spreads smaller doses over a longer period of time, allowing healthy cells to recover from less damaging single-strand breaks. Since cancer cells are more sensitive to ionizing radiation, they repair with less likelihood than do healthy cells.

Multi-leaf collimators are used to shape the beam like the tumor, spatially minimizing, as much as possible, the irradiation of healthy dimensions. To account for growth between planning and procedure, an assigned clinical target volume (CTV) is of greater radius than the initial tumor cross-section. In some procedures, molecular agents or brachytherapy may be used to lower the threshold of required radiation, delivering smaller amounts of damage to healthy tissue. 

\vspace{-3mm}
\subsection{Experimental Setup}

Geant4 software will be used to investigate the theoretical passage of the different particles through a phantom, which is here defined as a 30cm x 30cm x 50cm volumetric box consisting of many layers of water, designed to simulate the human body. A beam of varying energies (with particle number set to 1000) will be directed at the phantom along the Z-component of a
three-dimensional Cartesian plane. A detector will measure the number of counts against the spatial displacement, generating histograms to record particle activity. The different therapies will be tested and compared, with explanation as to which therapies are more or less useful, depending upon the environment and niche application.

\vspace{-1mm}
\section{Photon beams}
%\subsection{Theory and Application}

Photons (X-rays) are electromagnetic waves with no mass or charge \cite{suit} that lose energy by the photoelectric effect, Compton scattering, Rayleigh scattering, and photo-nuclear absorption  \cite{groom}. Compton effects contribute the most to photon interactions between 0.1 and 10 MeV while pair production dominates for higher energies \cite{gazis}. Clinical radiotherapies are delivered by a linear accelerator with energies ranging from 5 to 25 MeV. Compton scattered photons are provided to the tumor site by aligning maximal dose with the targeted region. Low surface-dose skin sparing is a characteristic benefit of photon therapy, but a slow exponential fall-off of radiation intensity after peak dose is usually more damaging than good. Long-range photon radiation better serves tumors that extend for longer lengths but that are not strictly confined to greater depths.

All radiotherapies vary with \textit{in vitro} and \textit{in vivo} environments. While it is always possible for radiation induced death in the former, it is not always possible in the latter, since the treatment of some localized cancer poses too much risk to healthy cells, hence establishing a critical clinical constraint: precision. Currently, the most precise method for photon beam therapy is intensity-modulated radiotherapy (IMRT) \cite{kooy}. Using multiple radiation fields of varying direction and intensity, the size of the clinical target volume (CTV) is increased. Since secondary radiation-induced malignancies occur ten to fifteen years after treatment, radiation exit dose is mostly considered a problem for pediatric patients.

\vspace{-1mm}
\subsection{Results and Discussion}

A 10 MeV photon beam was applied to the human-like phantom previously defined. Shown below are the histograms for the energy absorbed; the X, Y, and Z spatial components; and the energy deposited against Z.

Pair-production and Compton scattering dominate the photon energy loss at 10 MeV \cite{gazis}. Energy is also lost by the photoelectric effect and photo-nuclear absorption, but these losses are not observable for energies greater than an MeV \cite{groom}. It is particularly the Compton effect that delivers the absorbed dose to the the tumor site.

\vspace{-1mm}
\begin{figure}[H]
    \centering
    \includegraphics[width=8.5cm]{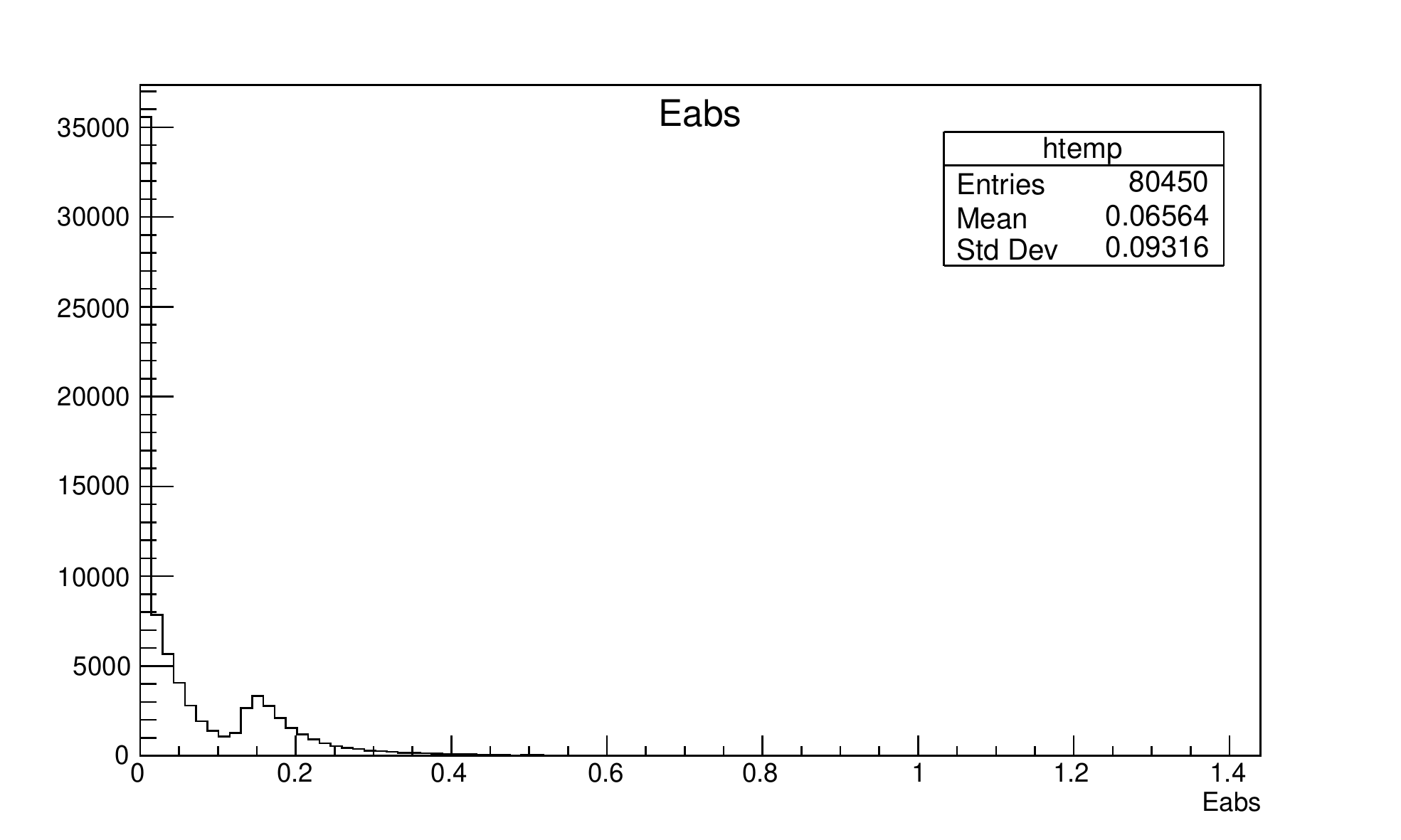}
    \caption{Energy absorbed histogram for photons with 10 MeV beam energy.}
    \label{fig:gammaeabs}
\end{figure}

\vspace{-1mm}
The energy absorbed histogram [Figure \ref{fig:gammaeabs}] shows Compton scattering with a peak. The peak may be scaled up or down as a function of the number of events by increasing or decreasing the beam energy, respectively. For the 10 MeV beam, the mean length of radiation interaction is 2.83282cm. It increases to 5.41134cm for a 20 MeV beam and decreases to 1.53814cm for a 5 MeV beam, in each case varying by a factor of roughly two. This informs the conclusion that the depth of peak radiation is directly dependent on beam energy, though not nearly by a factor of two; doubling the energy only shifts the distribution to slightly deeper regions.

\begin{figure}[h!]
    \centering
    \begin{minipage}{0.5\textwidth}
        \centering
        \includegraphics[width=8.5cm]{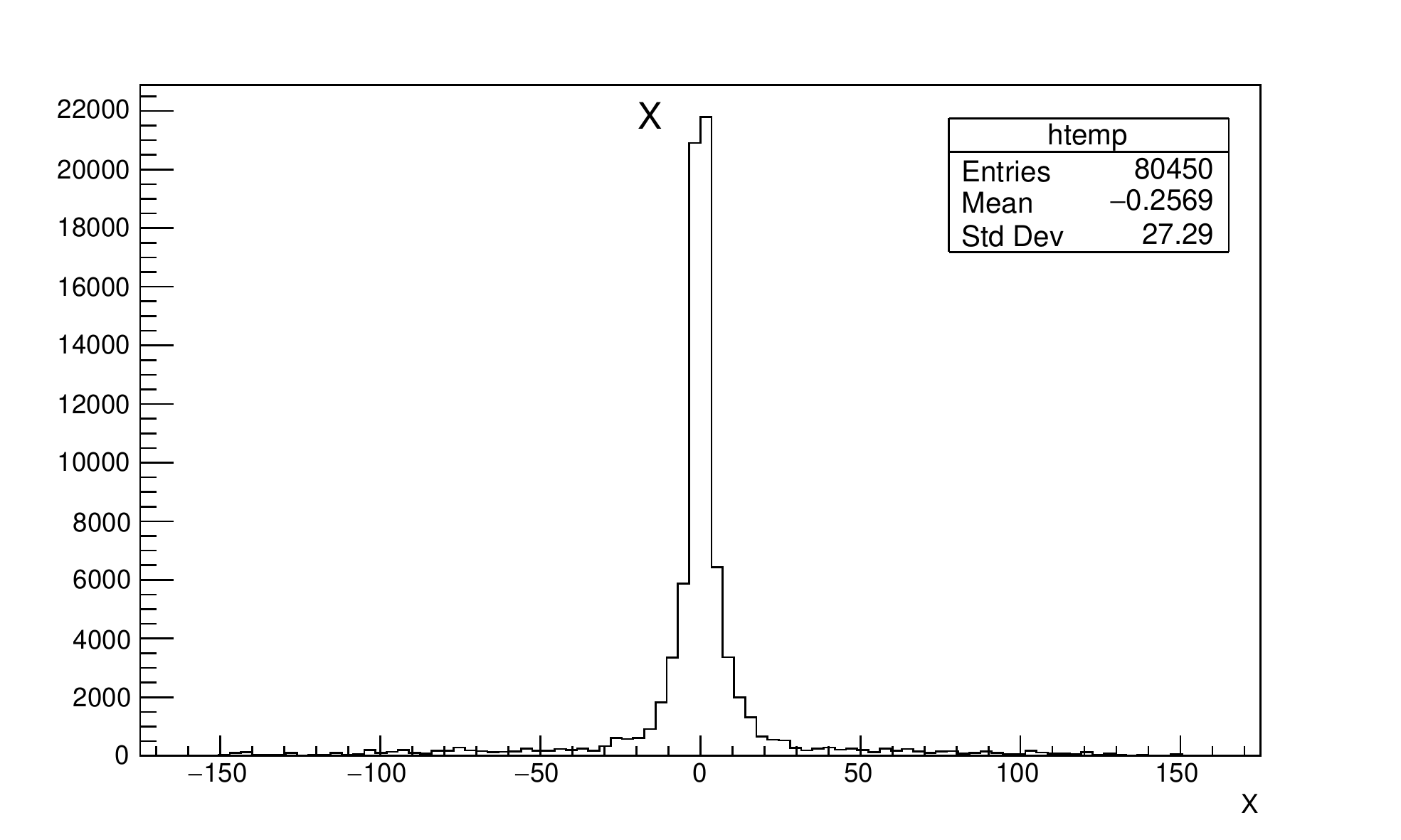}
        \caption{X-direction histogram.}
        \label{fig:gammax}
    \end{minipage}%
    \begin{minipage}{0.5\textwidth}
        \centering
        \includegraphics[width=8.5cm]{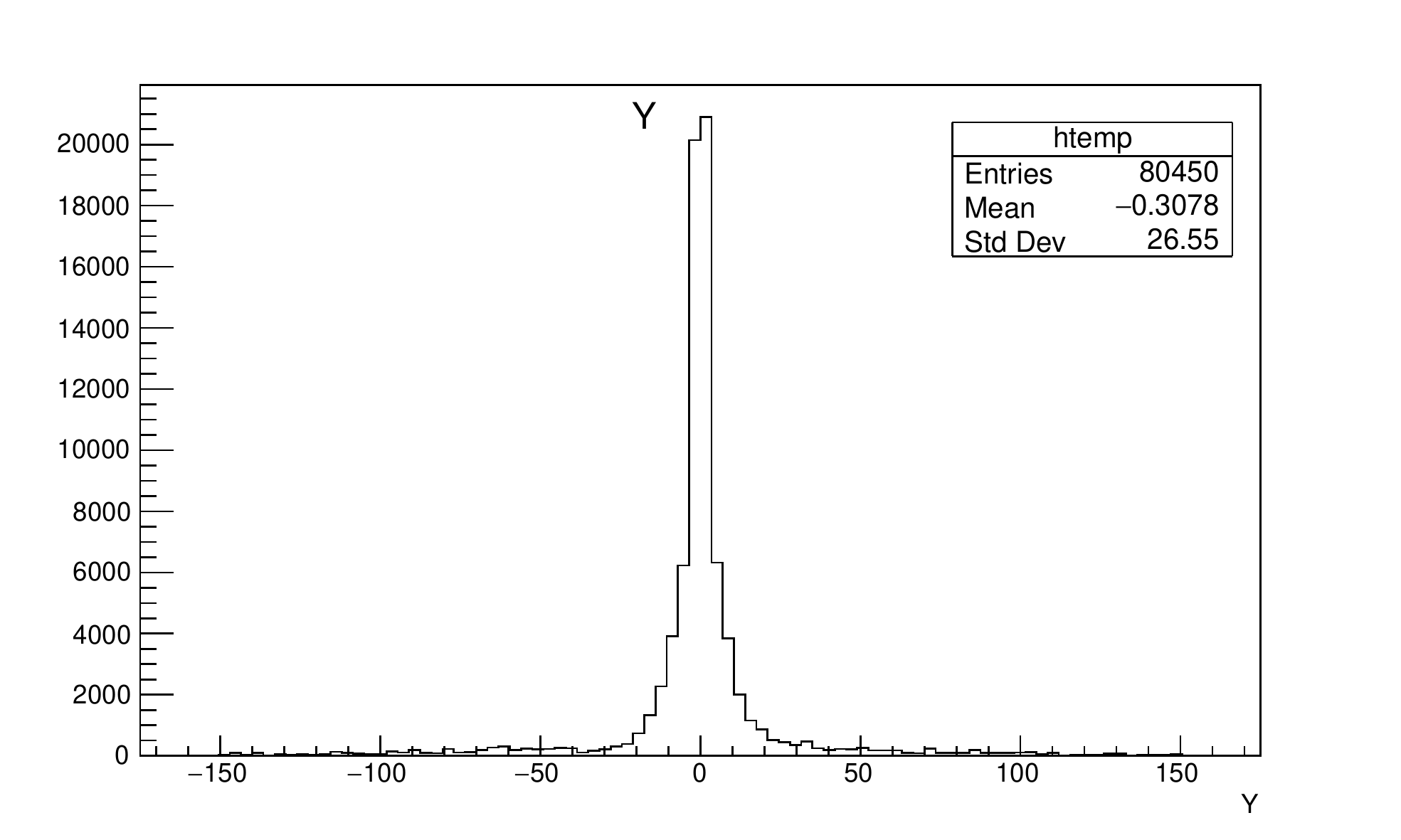}
        \caption{Y-direction histogram.}
        \label{fig:gammay}
    \end{minipage}
\end{figure}

\begin{figure}[h!]
    \centering
    \begin{minipage}{0.5\textwidth}
        \centering
        \includegraphics[width=8.5cm]{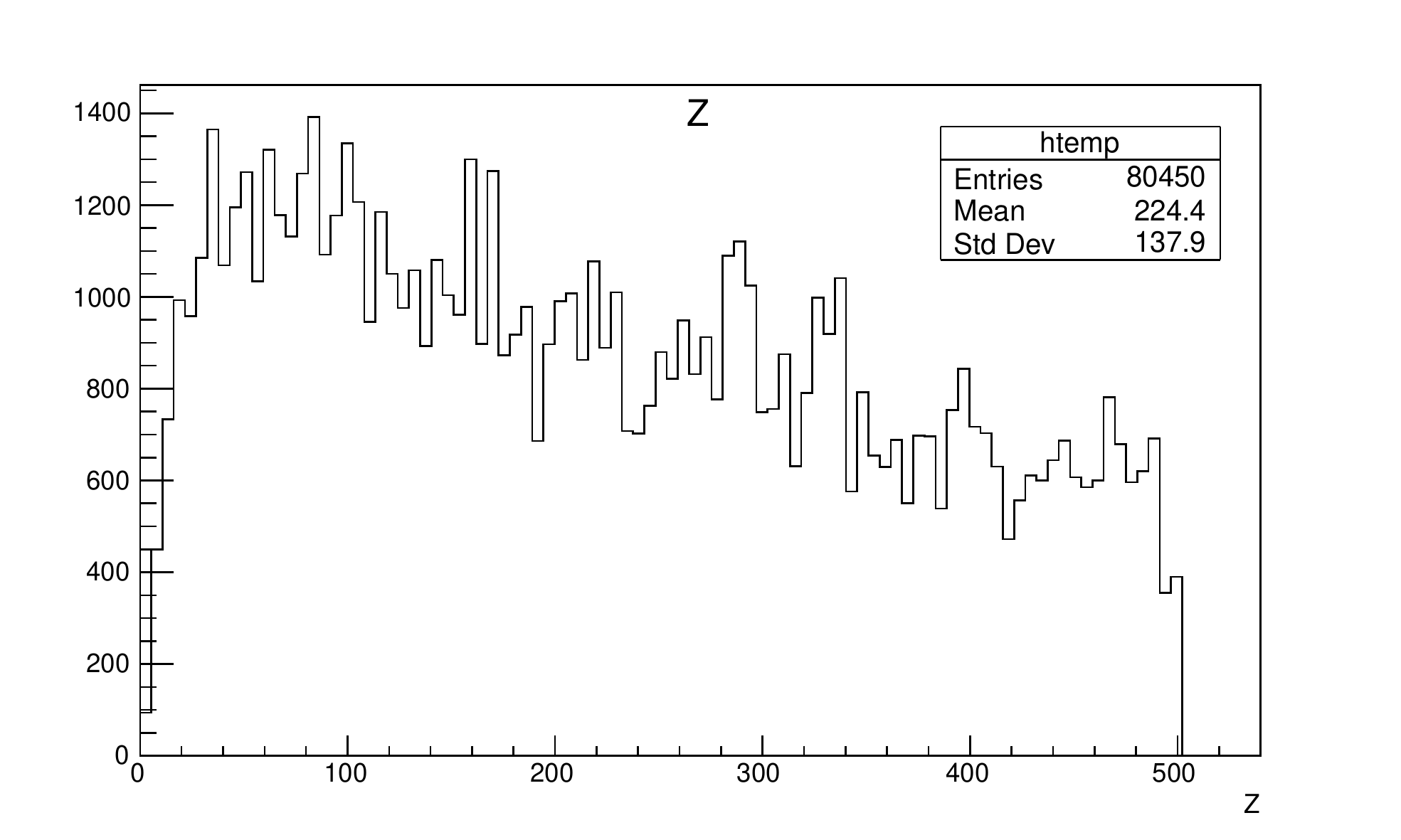}
        \caption{Z-direction histogram.}
        \label{fig:gammaz}
    \end{minipage}%
    \begin{minipage}{0.5\textwidth}
        \centering
        \includegraphics[width=8.5cm]{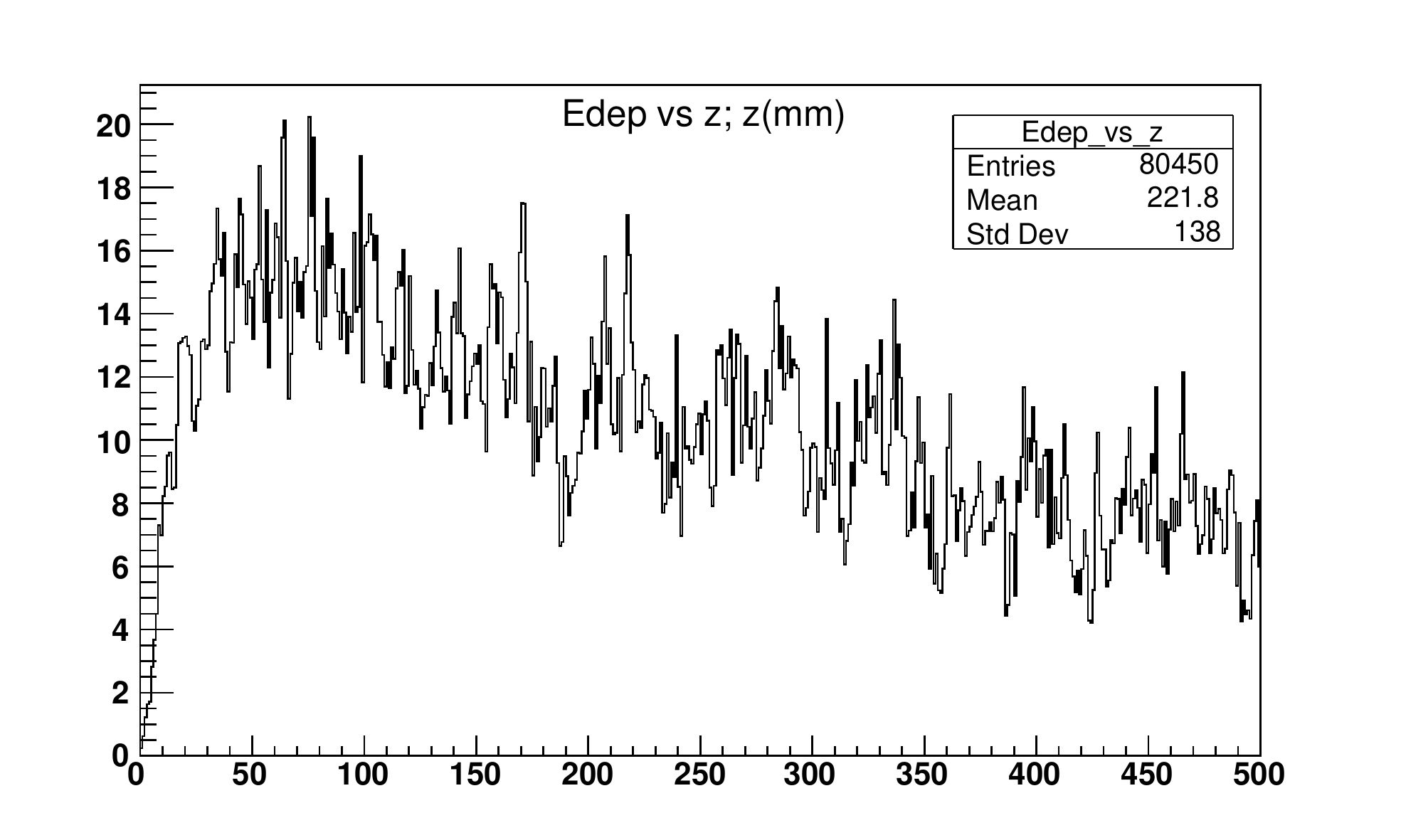}
        \caption{Energy deposited histogram.}
        \label{fig:gammaedep}
    \end{minipage}
\end{figure}

The X– and Y–direction histograms [Figures \ref{fig:gammax} and \ref{fig:gammay}] show particles scattering parallel to the phantom surface and perpendicular to the beam, as expected. The Z(beam)–direction histogram [Figure \ref{fig:gammaz}] shows a strong correlation between the number of events and the energy deposited. Likewise, Figure \ref{fig:gammaedep} shows initial skin sparring as the energy deposited rises to its 20 keV peak. 

It is observed that the maximum dose is delivered to the tumor site between 6.5cm and 7.5cm. This peak is followed by a long radiation tail for decreasing energy, with continuous fall-off throughout the phantom. The beam penetrates all 50cm of the \textit{patient} but not with all 10 MeV of energy deposited. A close order of 4.5 MeV is not detected. Since the energy decline is roughly linear, the total energy deposited after the peak may be approximated using a simple point-slope calculation: $E_{1} \approx 17$ keV at 7.5cm and $E_{2} \approx 6$ keV at 50cm. The line of descent has the form

\begin{equation*}
    E - E_{2} = \frac{E_{1} - E_{2}}{Z_1 - Z_2} (Z - Z_2),
\end{equation*}

\noindent which becomes
\begin{equation*}
    E = -0.026(Z - 500) + 6.
\end{equation*}
\noindent Assuming the maximum dose is delivered at the tumor site, the exit dose is then the integral dose of the energy deposited after the peak:

\begin{equation*}
    E_\text{exit} \approx \int_{Z_1}^{Z_2}E  \,dZ  \approx 4.9 \text{ MeV}. 
\end{equation*}

\noindent Using this crude approximation, and accounting for over– and under–estimations before and after the required dose, the amount of radiation actually provided to the target volume is roughly \textit{one-tenth} the total amount of radiation deposited into the patient.

\begin{figure}[H]
    \centering
    \includegraphics[width=8.5cm]{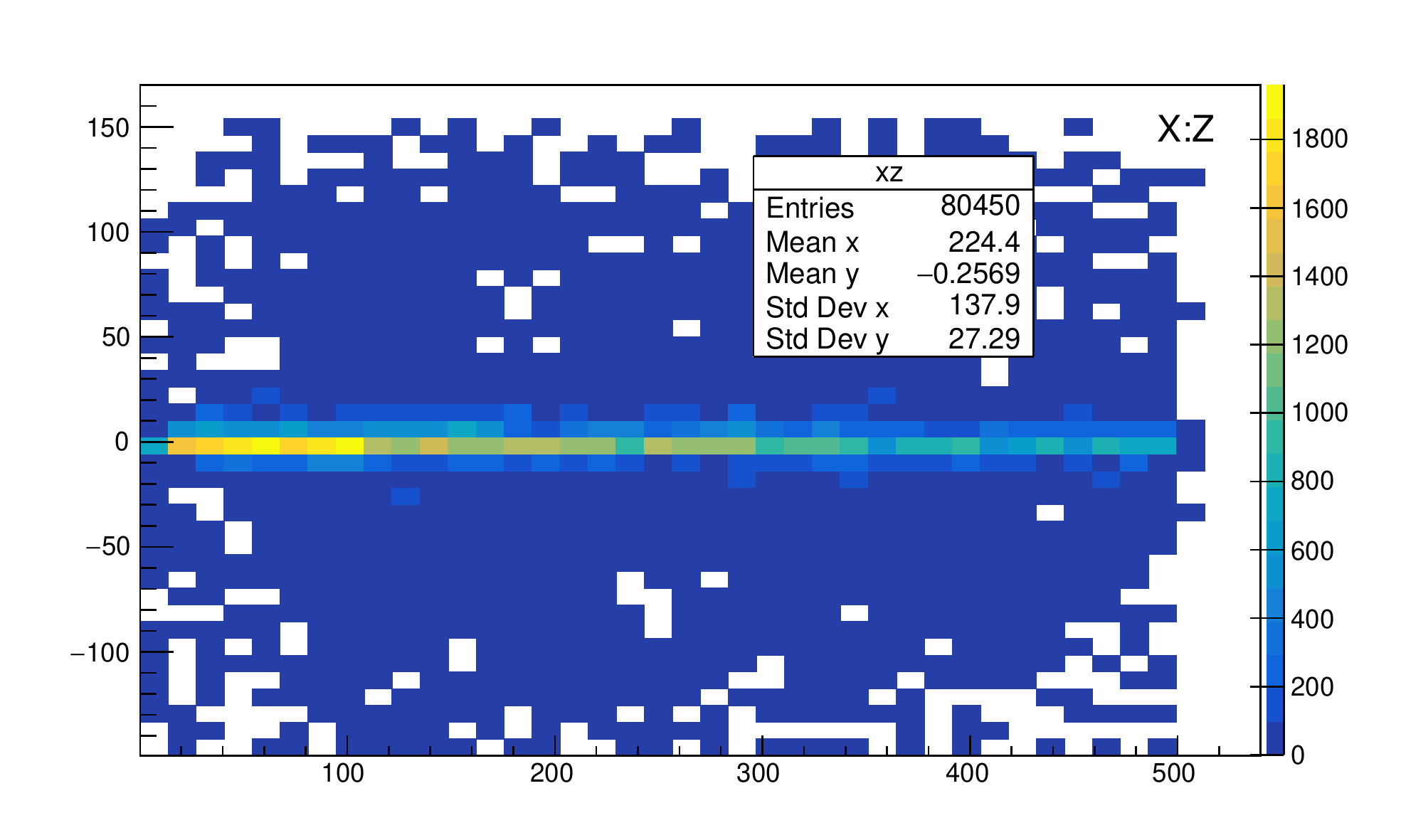}
    \caption{X:Z heat map for 10 MeV photons.}
    \label{fig:gamma10-heat}
\end{figure}

\begin{figure}[h!]
    \centering
    \begin{minipage}{0.5\textwidth}
        \centering
        \includegraphics[width=8.5cm]{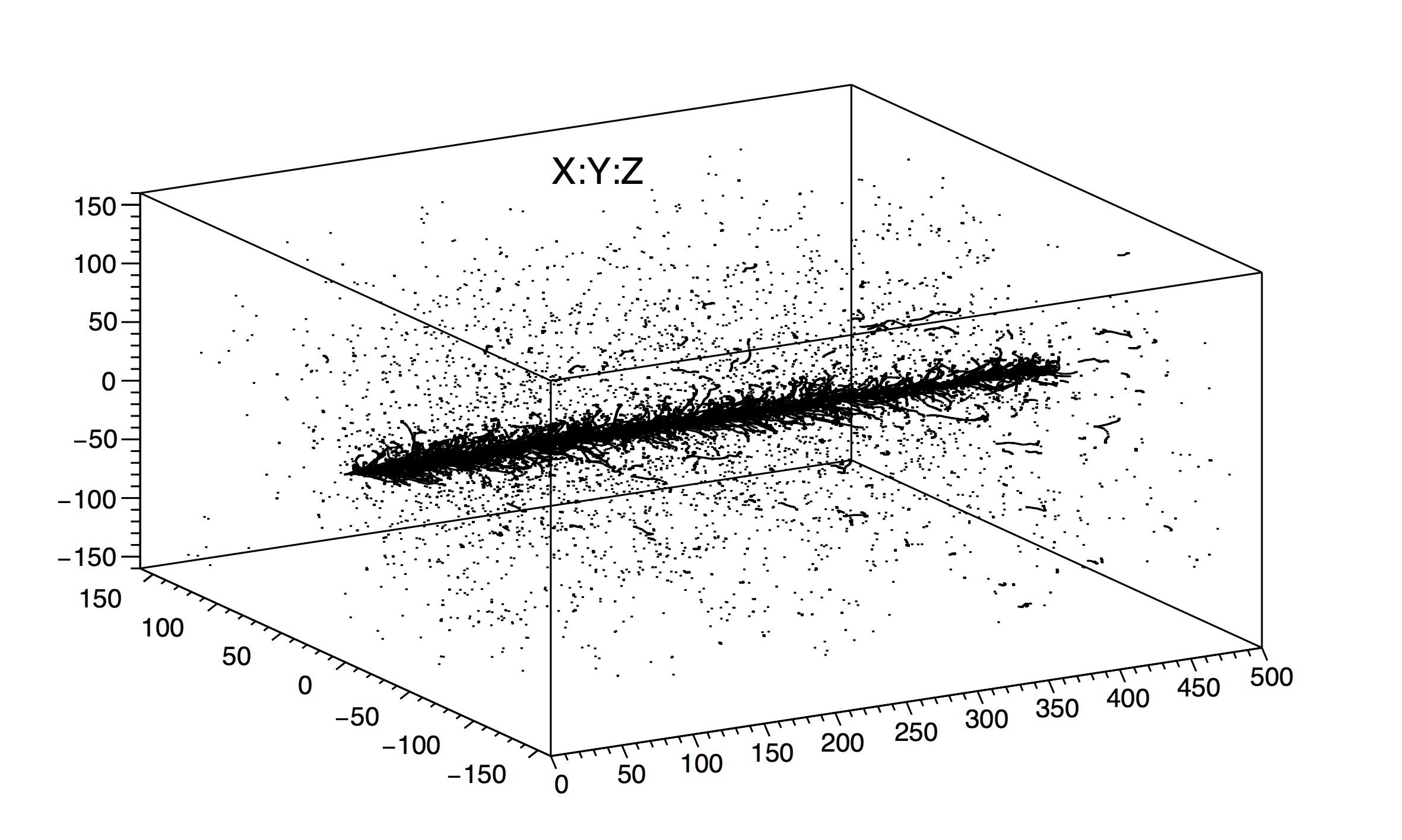}
        \caption{X:Y:Z for 10 MeV.}
        \label{fig:gamma10-3D}
    \end{minipage}%
    \begin{minipage}{0.5\textwidth}
        \centering
        \includegraphics[width=8.5cm]{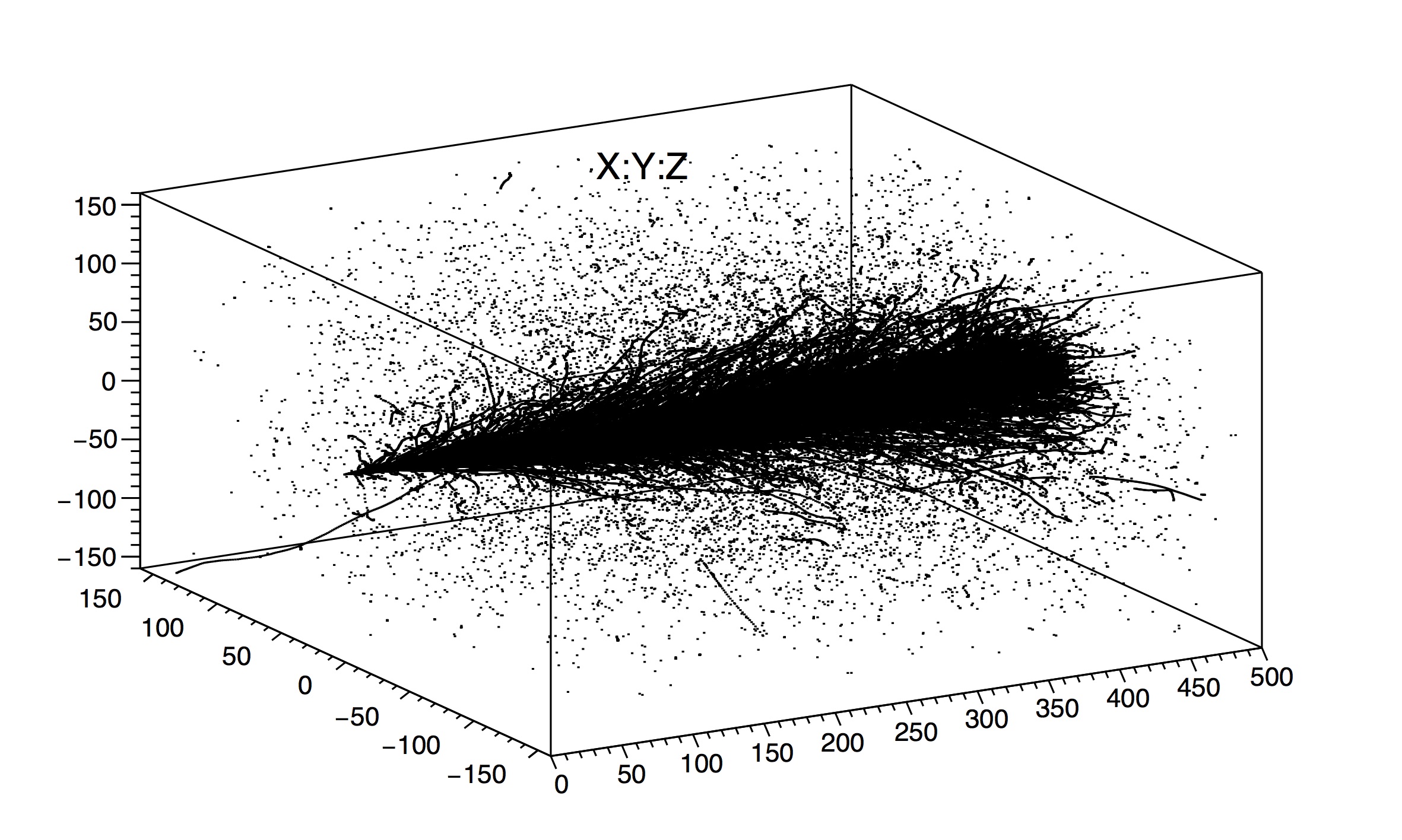}
        \caption{X:Y:Z for 100 MeV.}
        \label{fig:gamma100-3D}
    \end{minipage}
\end{figure}

The Z-track and intensity of beam may be observed by the heat map [Figure \ref{fig:gamma10-heat}]. The beam maintains consistent intensity throughout the entire system, with exponential fall-off beyond peak delivery.
\vspace{-6mm}
\vspace{-3mm}
To see the difference in radial profile for higher energies, three-dimensional displacement graphs were computed for 10 MeV and 100 MeV photon beams [Figures \ref{fig:gamma10-3D} and \ref{fig:gamma100-3D}]. In comparison to the 10 MeV case, the 100 MeV beam delivered its maximum dose near the boundary of the 50cm-wide phantom, with more divergence in the X– and Y–directions. It may then be inferred that for photons the radial dose increases with energy as does the depth of maximum dose. This imposes depth constraints for deeper tumors, where surrounding tissue may be adversely exposed to unintended ionizing radiation.
\vspace{-1mm}

\section{Proton Beams}

%\subsection{Theory and Application}

Unlike photons, protons are heavy charged particles with finite penetration depth and a characteristic Bragg peak (where maximum energy is deposited) \cite{kooy}. The relative biological effectiveness (RBF) of protons is roughly the same as photons, though their physical dose distributions differ sharply. Proton energy is lost immediately after the Bragg peak, giving negligible exit dose \cite{neutrons}. Since the direction of fired protons is changed when the particles are suddenly slowed by the Coulomb interaction, a wider dose at the Bragg peak is achieved for more elastic scattering events.

\noindent Skull sarcomas and eye melanomas have achieved the most clinical attention using proton therapies since these sites are often surrounded by critical anatomy, such as optic nerves and brain tissue \cite{kooy}. Delivered via a cyclotron, only a single proton field of varying energy is required to permeate the target volume. Hence, it is often the technical limitations on field size that limit the depth and volume of treatment. With energies varying between 70 and 250 MeV, some hospital cyclotrons have achieved accurate treatment up to 30cm of depth.

Proton therapy is more expensive than photon therapy by about a factor of 2.5 (IMRT), but is comparable to some brachytherapy costs (such as plaque) \cite{cost}. Proton therapy is made more difficult by a lack of modern imaging modalities, which require that larger margins of error be considered. Only a few proton facilities in the world have advanced volumetric imaging, whereas cone-beam CT imaging has been used in linear accelerators for the last twenty years.

\subsection{Results and Discussion}

\noindent A 100 MeV proton beam was applied to the human-like phantom. Shown below are the histograms for the energy absorbed; the X, Y, and Z spatial components; and the energy deposited against Z.

\begin{figure}[H]
    \centering
    \includegraphics[width=8.5cm]{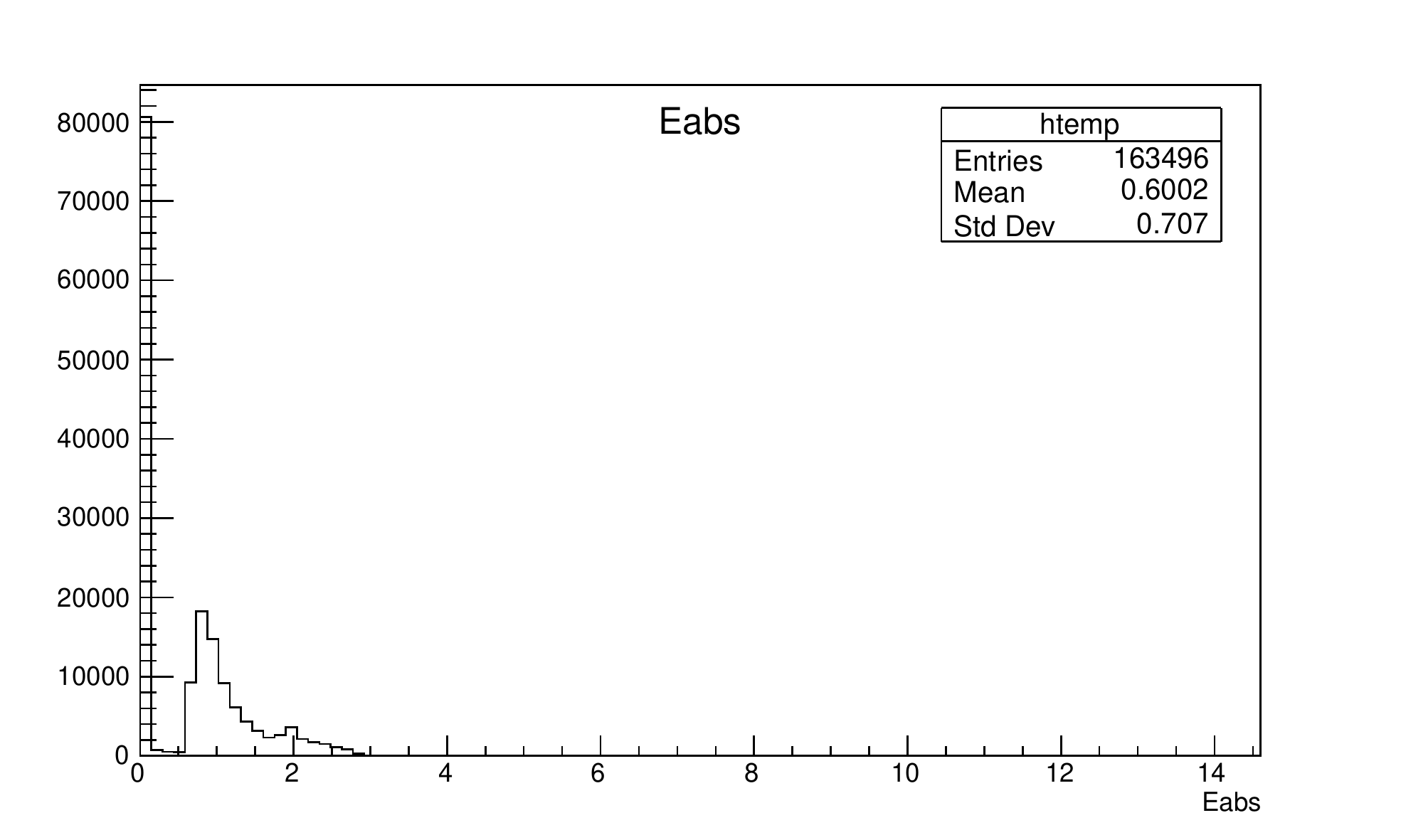}
    \caption{Energy absorbed histogram for protons with 100 MeV beam energy.}
    \label{fig:protoneabs}
\end{figure}

\begin{figure}[h!]
    \centering
    \begin{minipage}{0.5\textwidth}
        \centering
        \includegraphics[width=8.5cm]{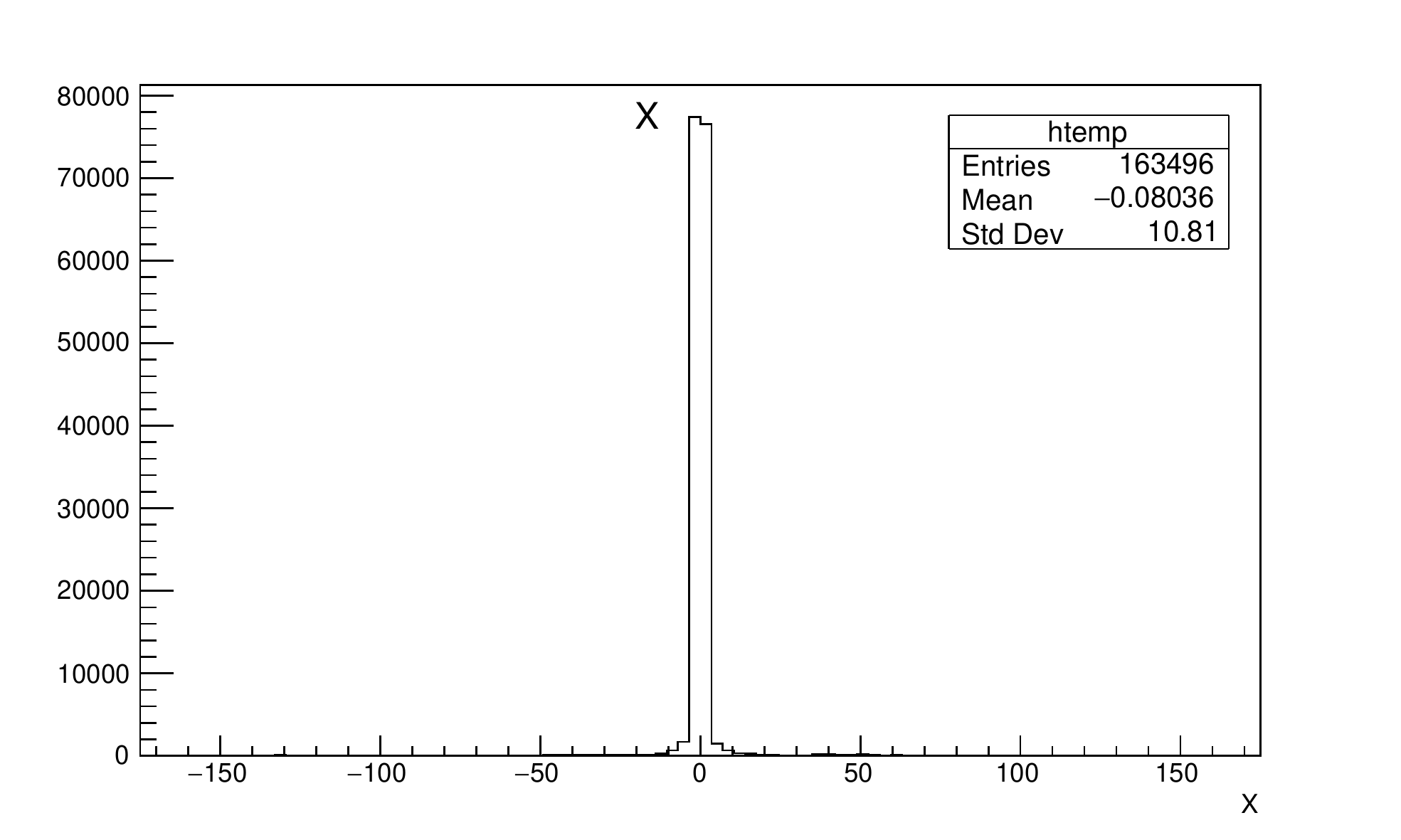}
        \caption{X-direction histogram.}
        \label{fig:protonx}
    \end{minipage}%
    \begin{minipage}{0.5\textwidth}
        \centering
        \includegraphics[width=8.5cm]{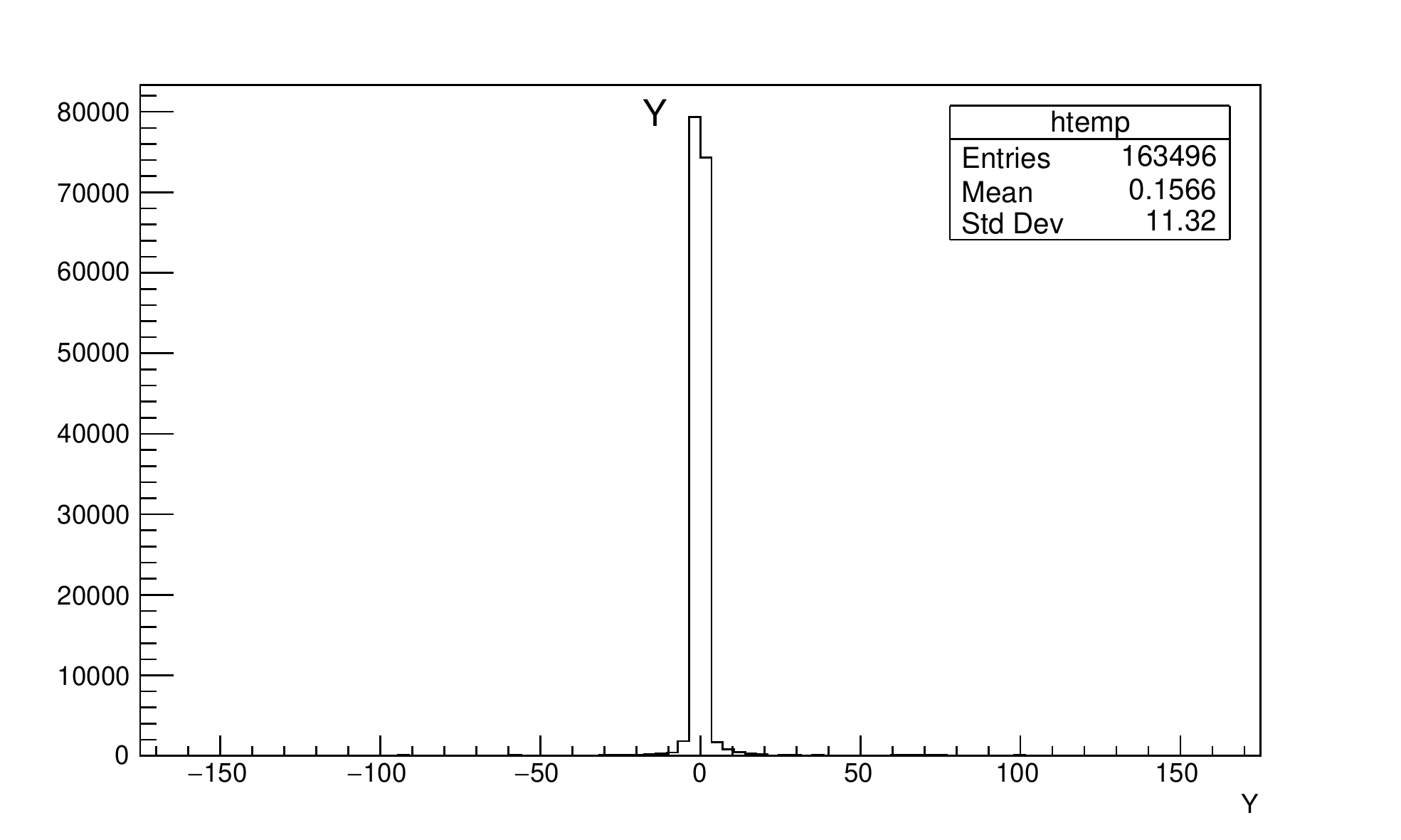}
        \caption{Y-direction histogram.}
        \label{fig:protony}
    \end{minipage}
\end{figure}

Figures \ref{fig:protoneabs}, \ref{fig:protonx}, and \ref{fig:protony} show similar characteristics to photons, owing similarities to radial profile and a high number of initial counts followed by a peak. 
\\
\\

\begin{figure}[h!]
    \centering
    \begin{minipage}{0.5\textwidth}
        \centering
        \includegraphics[width=8.5cm]{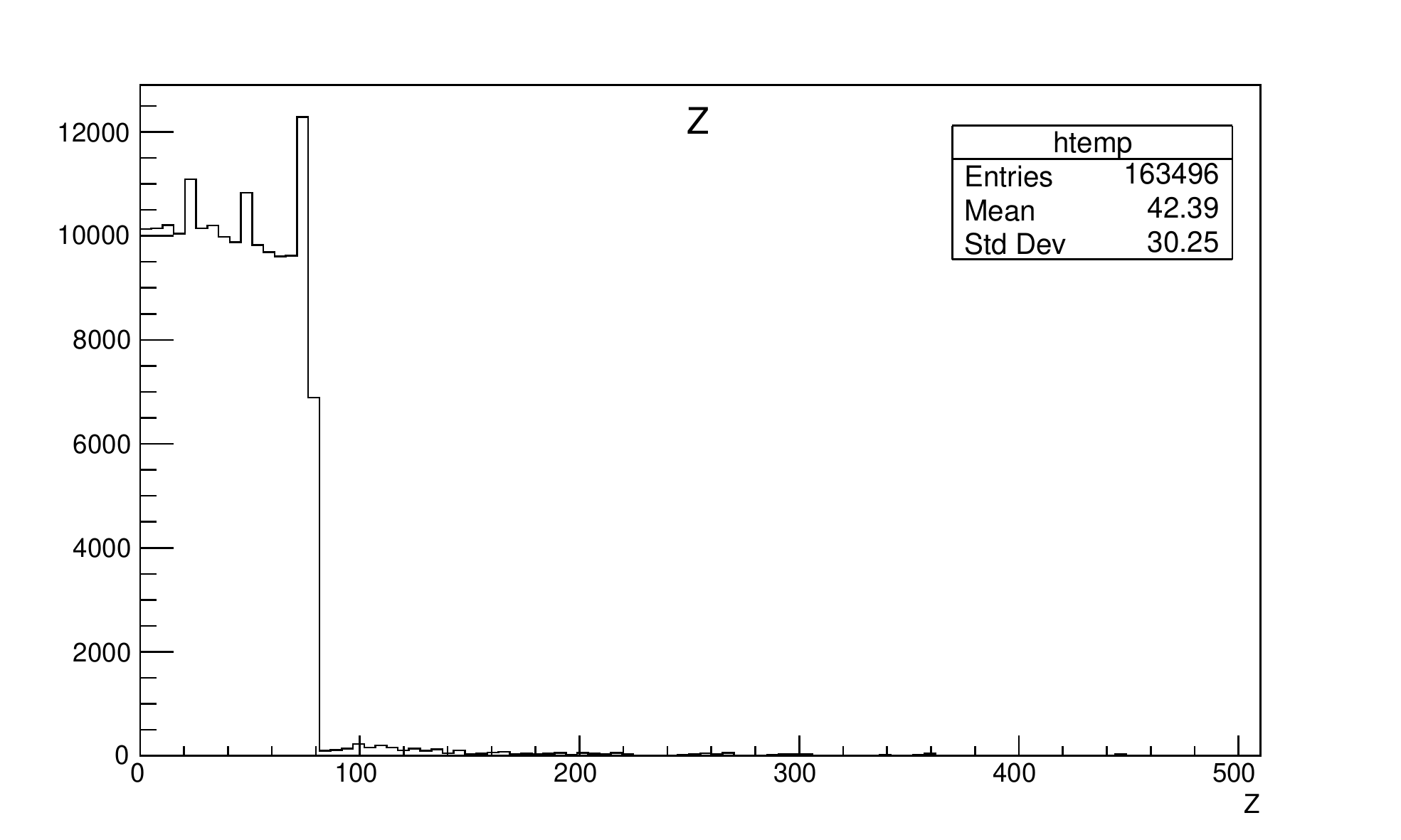}
        \caption{Z-direction histogram.}
        \label{fig:protonz}
    \end{minipage}%
    \begin{minipage}{0.5\textwidth}
        \centering
        \includegraphics[width=8.5cm]{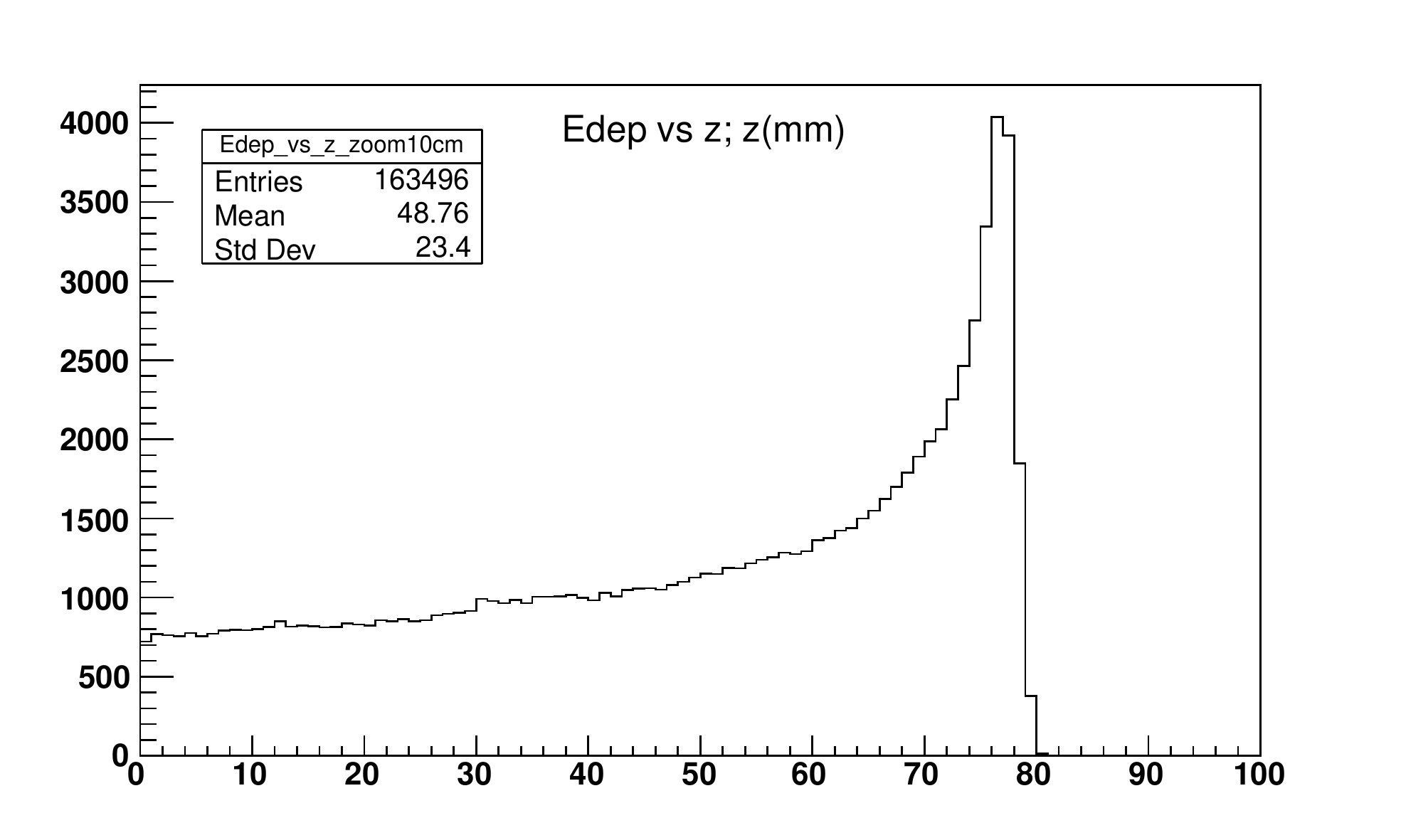}
        \caption{Energy deposited histogram.}
        \label{fig:protonedep}
    \end{minipage}
\end{figure}

The Z–direction histogram [Figure \ref{fig:protonz}] again shows strong correlation between the count number and the energy deposited [Figure \ref{fig:protonedep}]. At around 77mm the energy immediately falls-off, as does the number of counts measured in the detector. At the beginning of their path, the protons have high energy with a low elastic scattering cross-section due to proton absorption and electron scattering \cite{neutrons}. Most of the energy is then deposited at the Bragg peak, since there is a larger elastic scattering cross-section of low energy protons (due to a greater amount of Coulomb scattering). The increased number of elastic events changes the direction of the protons, giving a wider dose at the peak, which for this case is about 3mm.
\vspace*{-1mm}

\noindent The total energy deposited is

\begin{equation*}
    E_\text{dep} = E_\text{entry} +E_\text{target} + E_\text{exit}.
\end{equation*}

\noindent The radiation delivered to the target volume may be under-approximated by the width of the Bragg peak times the height, which for 100 MeV protons (with beam number set to 1000) is
\\
\vspace*{2mm}
\begin{equation*}
    E_\text{target} \approx (77\text{mm}-74\text{mm})\left(\frac{4000 \text{ counts}}{1000 \text{ p$^+$}}\right) = 12 \text{ MeV}. 
\end{equation*}
\\
\noindent Then assuming negligible exit dose, around 12\% of the received dose is provided to the target volume (compared to 10\% for photons).

\begin{figure}[h!]
    \centering
    \begin{minipage}{0.5\textwidth}
        \centering
        \includegraphics[width=8.5cm]{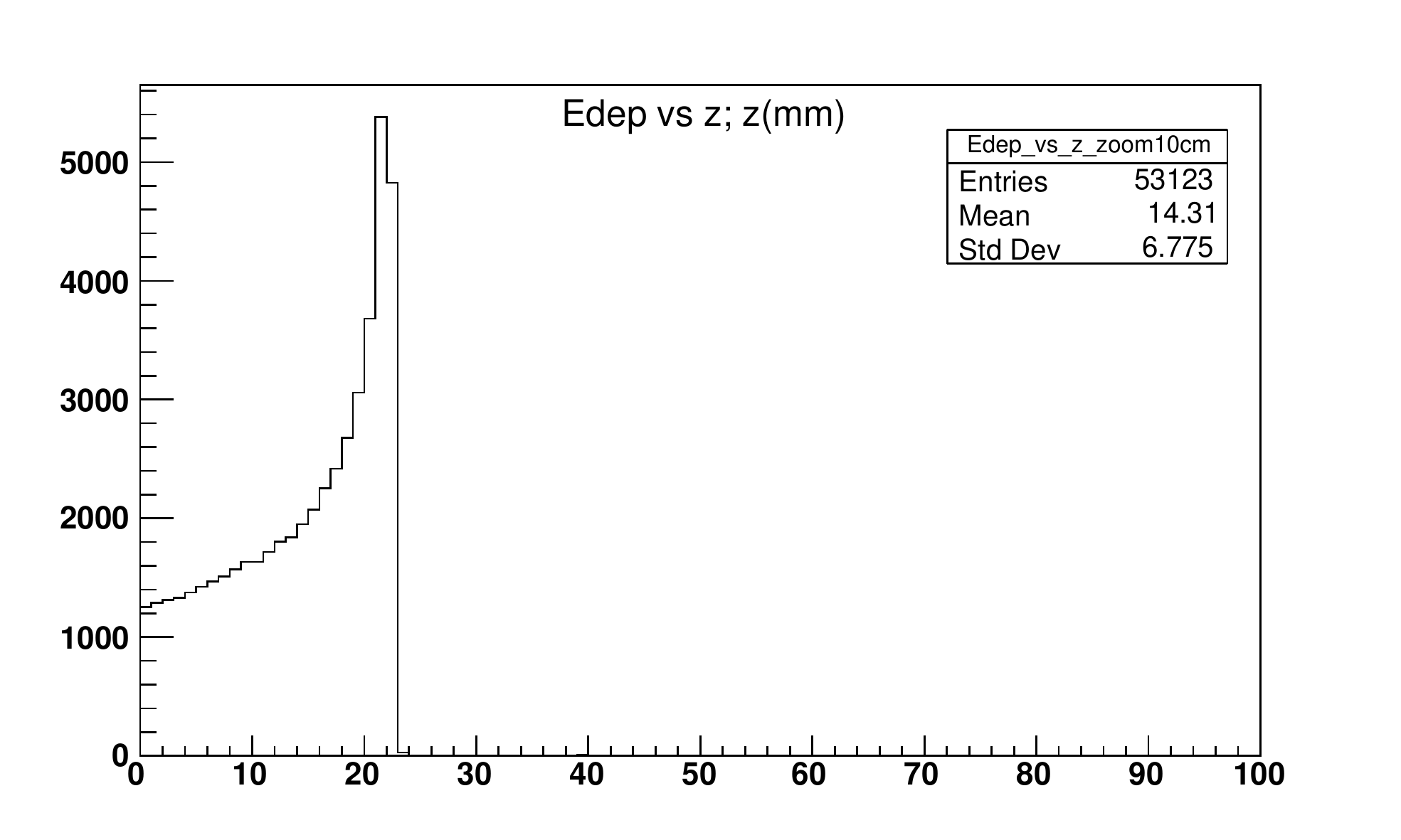}
        \caption{Edep vs Z for 50 MeV.}
        \label{fig:50proton}
    \end{minipage}%
    \begin{minipage}{0.5\textwidth}
        \centering
        \includegraphics[width=8.5cm]{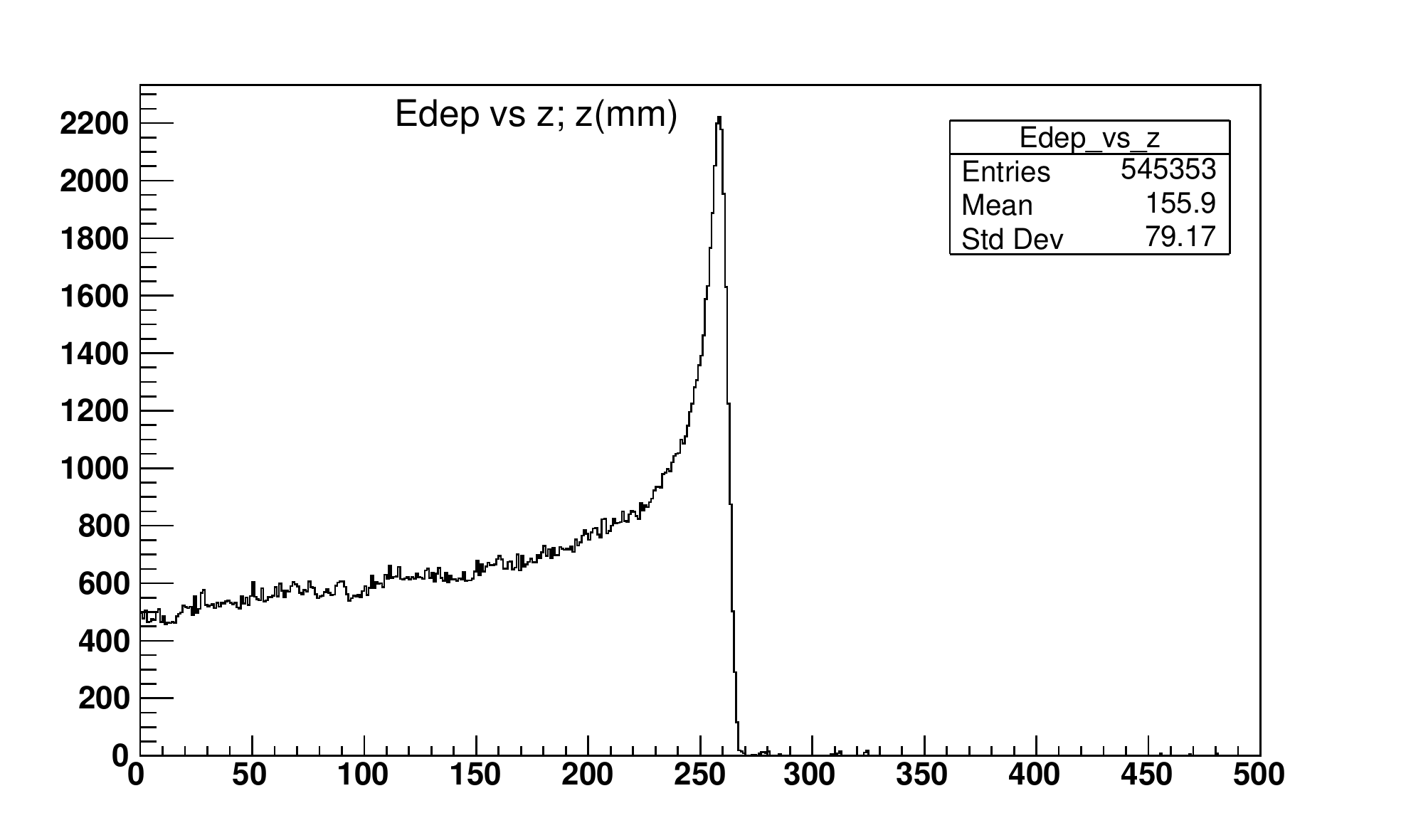}
        \caption{Edep vs Z for 200 MeV.}
        \label{fig:200proton}
    \end{minipage}
\end{figure}

For smaller and larger beam energies, a direct relationship between energy deposited and penetration depth is observed [Figures \ref{fig:50proton} and \ref{fig:200proton}]. For energies of 50, 100, and 200 MeV, the Bragg peak depth varies at roughly 21.5, 77, and 259mm, respectively. The width of the peak varies, too, with 200 MeV protons covering almost 9mm compared to 1.5mm for 50 MeV protons.

\vspace*{-5mm}

\begin{figure}[H]
    \centering
    \includegraphics[width=8.5cm]{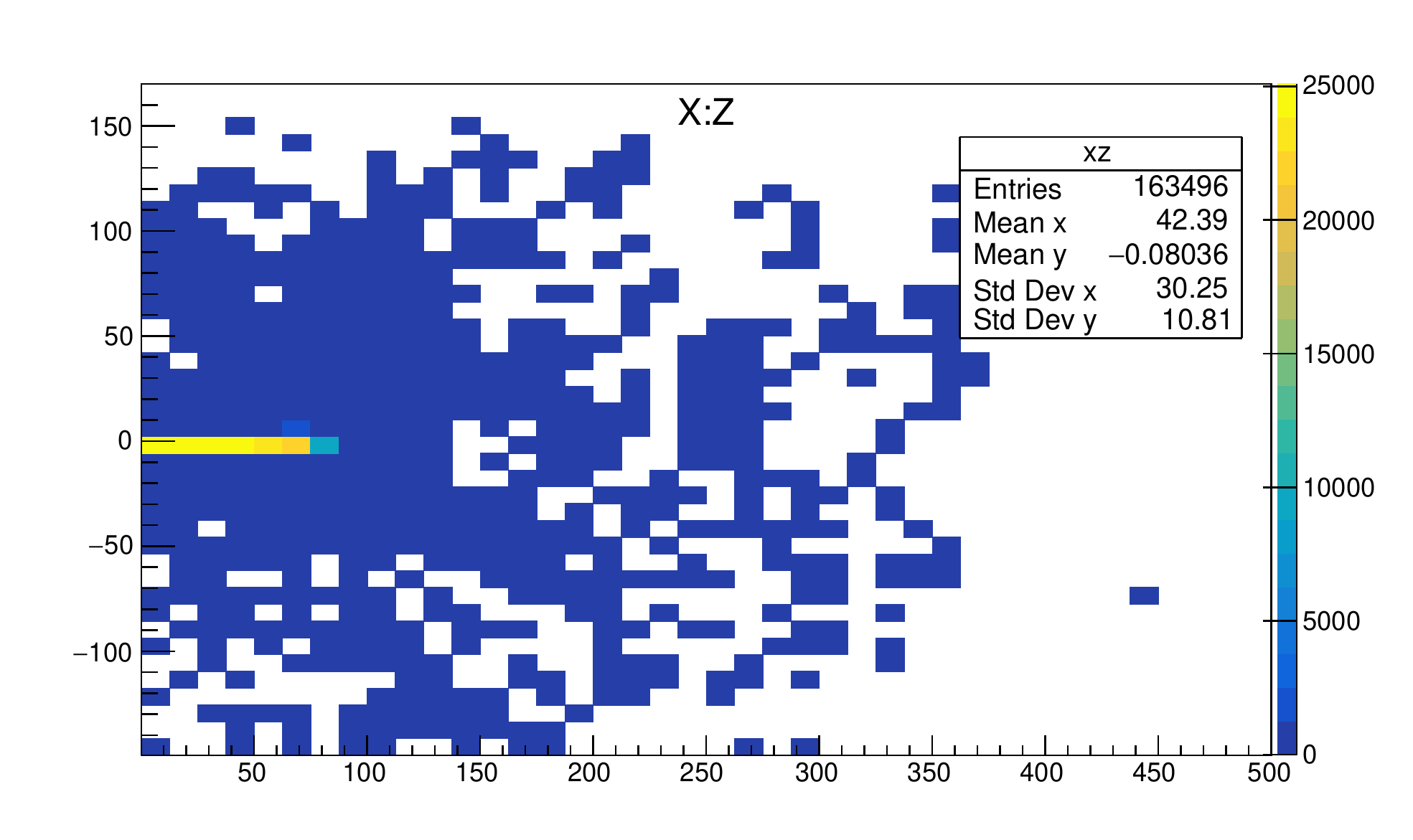}
    \caption{X:Z heat map for 100 MeV protons.}
    \label{fig:proton100-heat}
\end{figure}

\begin{figure}[h!]
    \centering
    \begin{minipage}{0.5\textwidth}
        \centering
        \includegraphics[width=8.5cm]{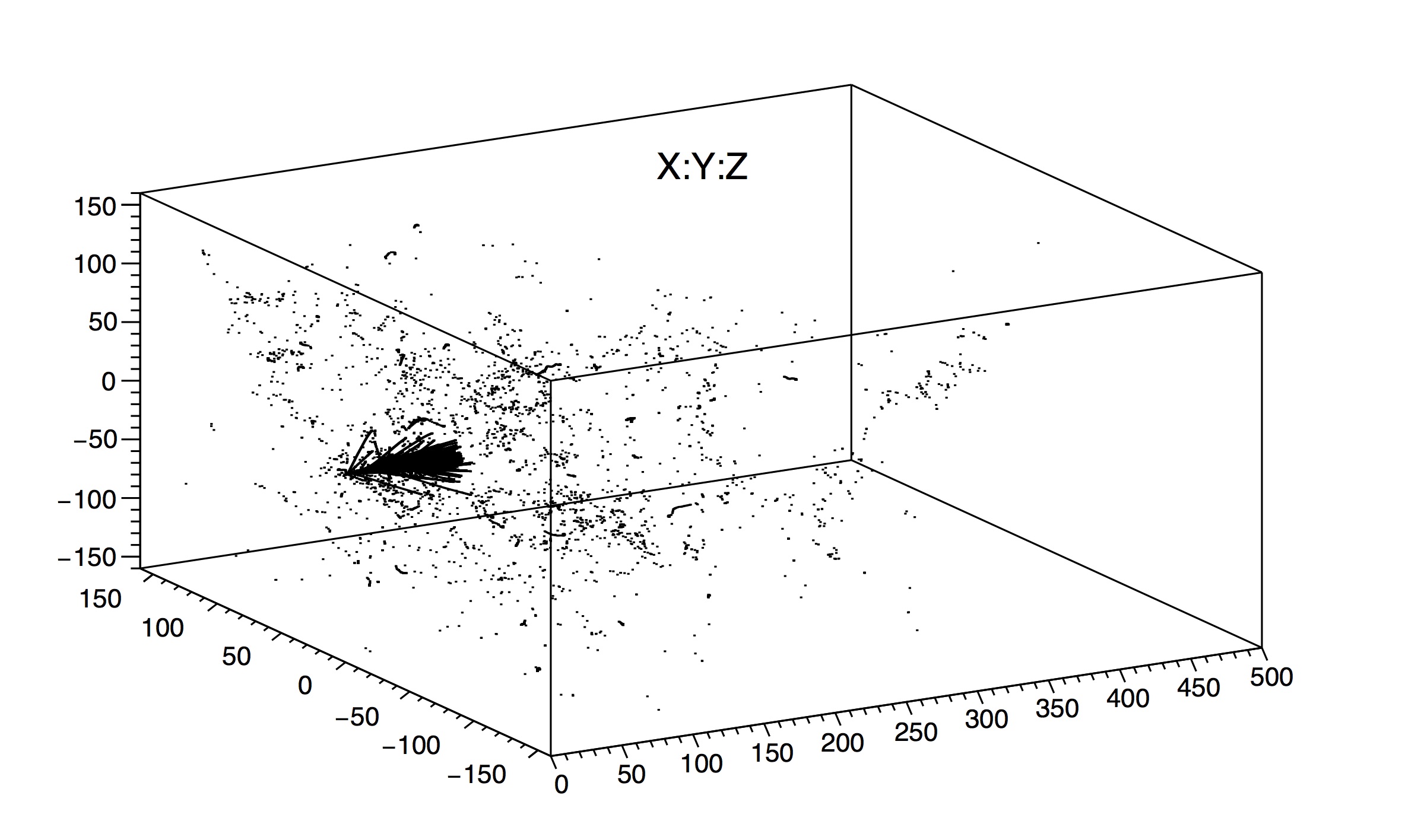}
        \caption{X:Y:Z for 100 MeV protons.}
        \label{fig:proton100-3D}
    \end{minipage}%
    \begin{minipage}{0.5\textwidth}
        \centering
        \includegraphics[width=8.5cm]{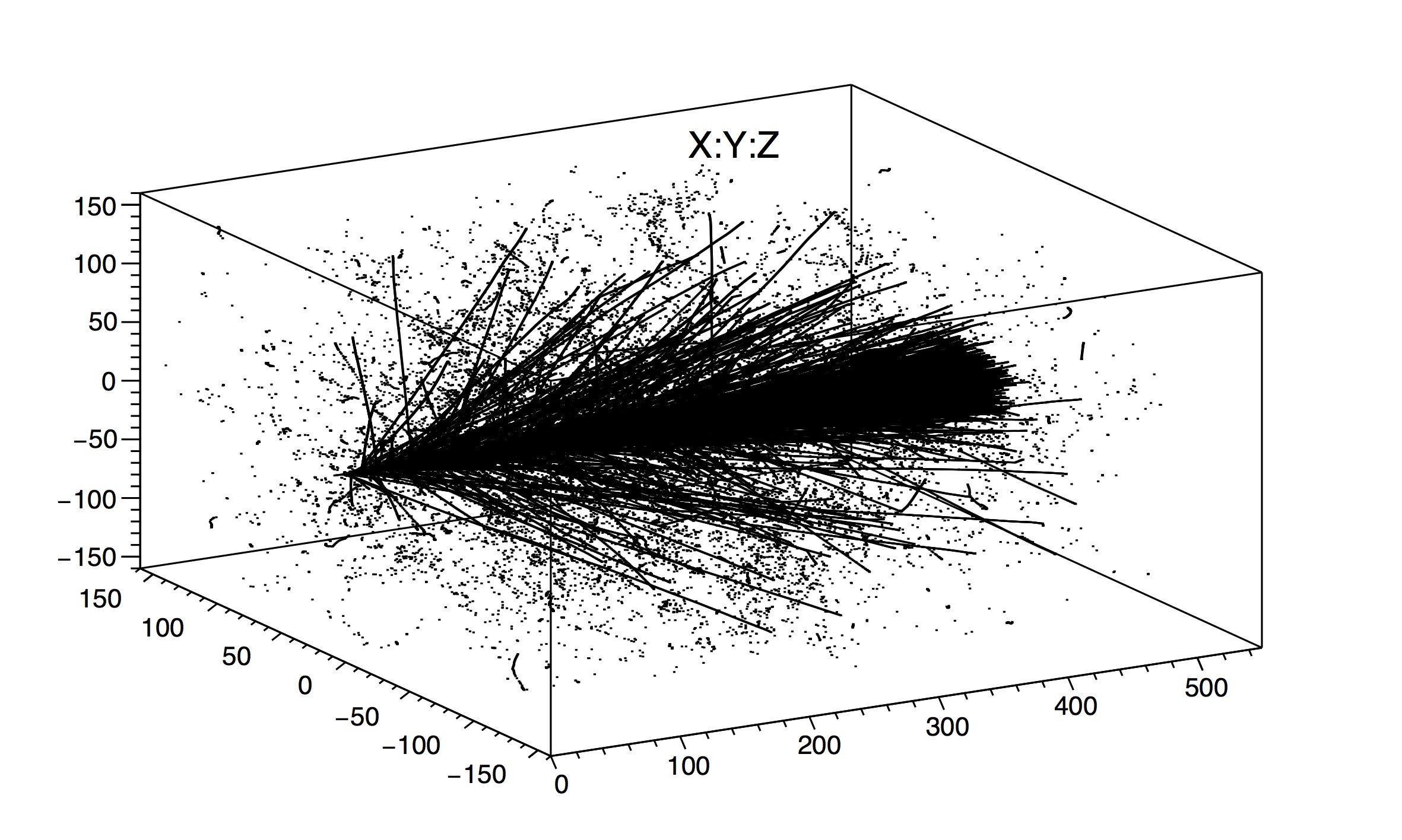}
        \caption{X:Y:Z for 290 MeV protons.}
        \label{fig:proton290-3D}
    \end{minipage}
\end{figure}

In comparison to 10 MeV photons, 100 MeV protons have limited range and are more localized [Figures \ref{fig:proton100-heat} and \ref{fig:proton100-3D}]. For greater energy, however, the radial profile expands spatially in the X– and Y–directions. Figure \ref{fig:proton290-3D} shows the extreme case of this, where for 290 MeV protons (corresponding to a maximum dose delivered at the far border of the phantom) the peripheral radiation dose is quite destructive. 

\vspace*{-3mm}

\vspace*{-8mm}

\newpage
It should be noted that the simple modelling does not reflect properly the amount of background dose administered over multiple treatment sessions. This contribution can be quite large since the Bragg peak must be widened by more fields in order to fill the entire tumor volume. Nonetheless, the results do support niche applications of protons to skull sarcomas and eye melanomas, since the exit dose is nearly absent.

\vspace*{-4mm}
\section{Further Analysis}
\begin{figure}[h!]
    \centering
    \begin{minipage}{0.5\textwidth}
        \centering
        \includegraphics[width=8.5cm]{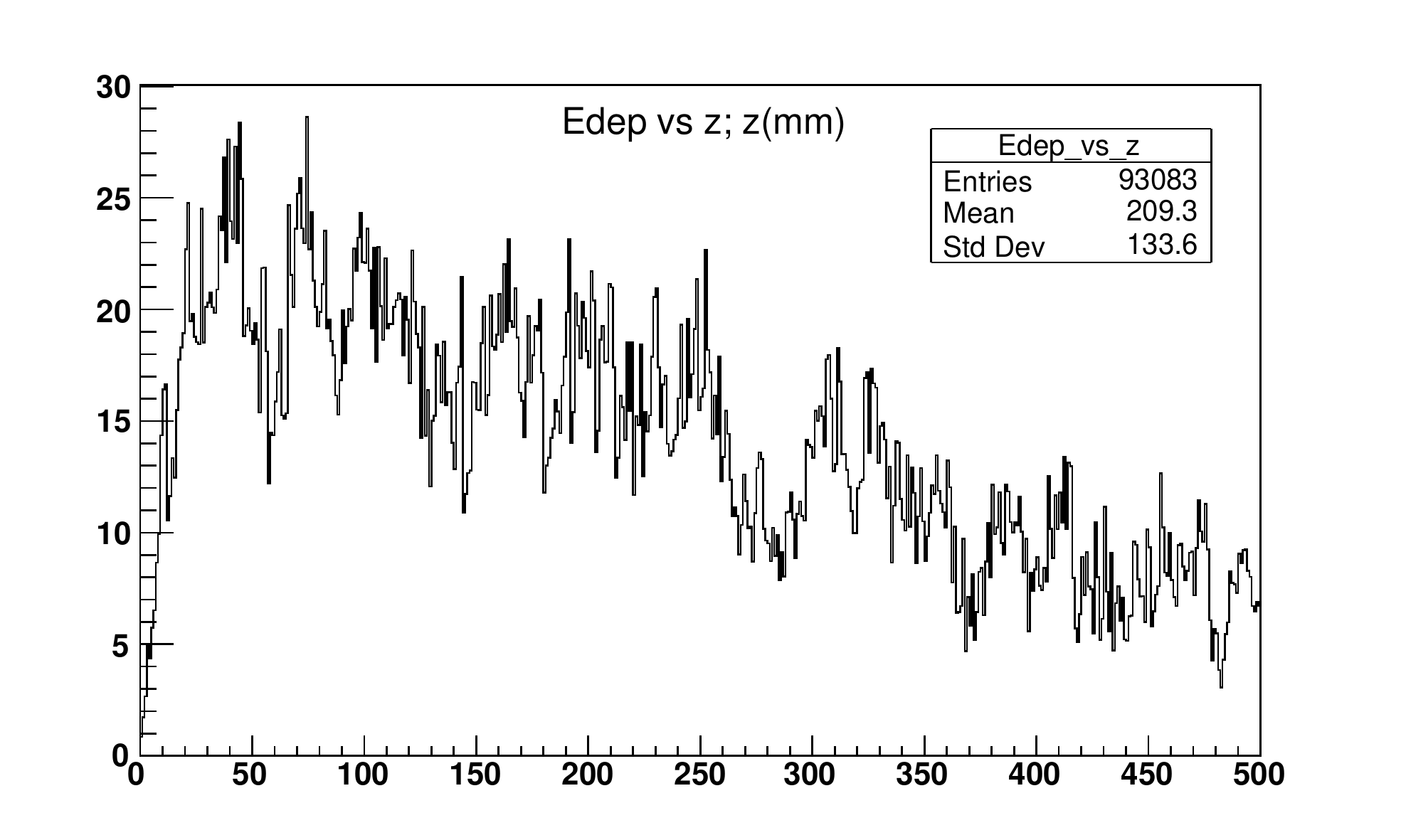}
        \caption{10 MeV photons in bone.}
        \label{fig:bone1}
    \end{minipage}%
    \begin{minipage}{0.5\textwidth}
        \centering
        \includegraphics[width=8.5cm]{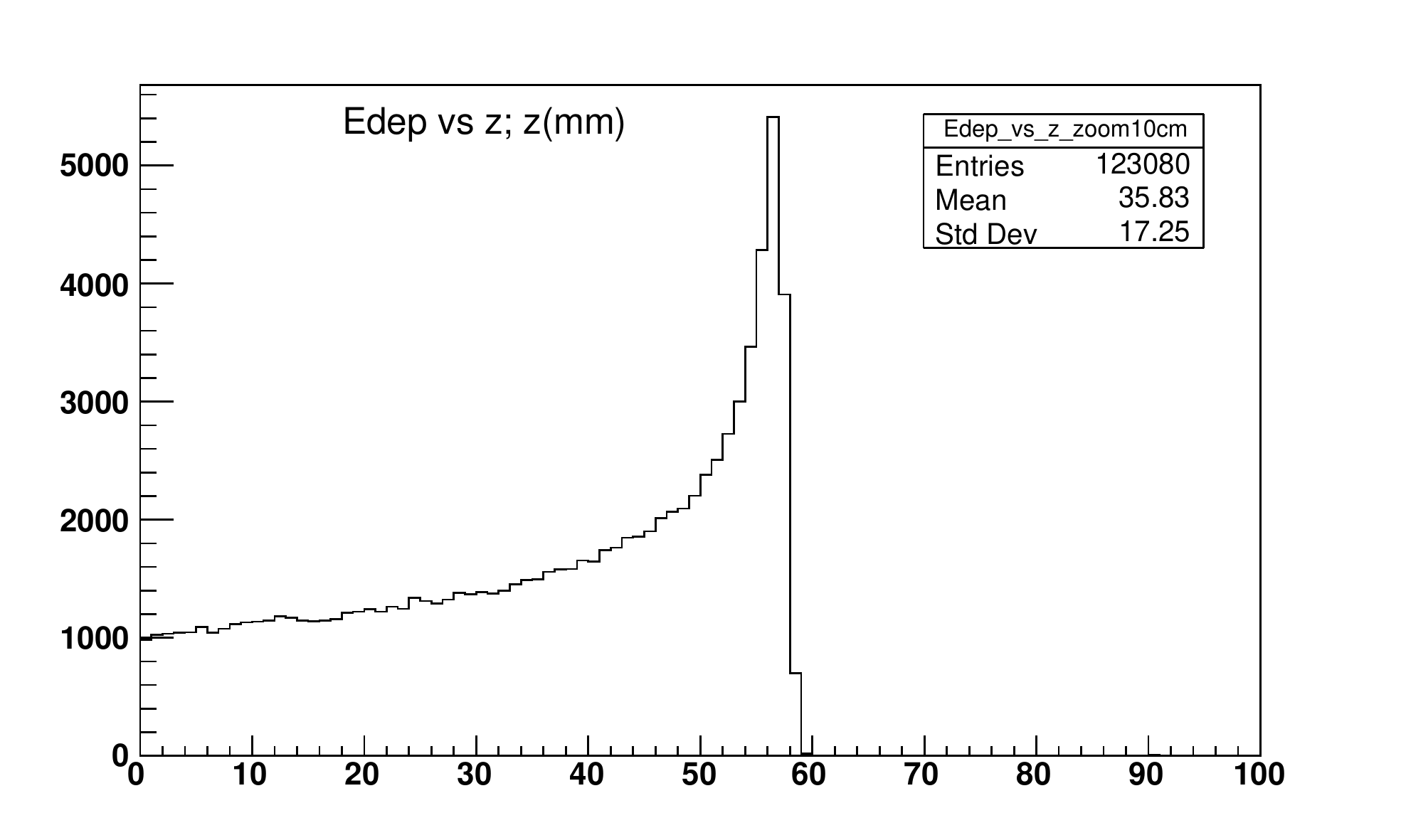}
        \caption{100 MeV protons in bone.}
        \label{fig:bone2}
    \end{minipage}
\end{figure}

\vspace*{-2mm}
For both photon and proton therapies it remains preferential to decrease target volume to limit the potential for secondary cancer. Since bone structure lies inherently in the path of many radiotherapies, the phantom was reconstructed using bone material in place of water. The energy deposited against Z was modelled for 10 MeV photons and 100 MeV protons, respectively [Figures \ref{fig:bone1} and \ref{fig:bone2}].

\vspace{-3mm}

The dose distribution for bone material changes less than might be expected, since solid bone and liquid water have surprisingly similar electron densities. The one noticeable difference was that slightly more dose was delivered in bone due to beam divergence and absorption within the heavier solid. The result is particularly important for photons, since the distinctively long radiation track will likely cross bone structures when exiting the body, administering a higher out-of-field dose.

It is uncommon to develop secondary cancer after radiotherapy treatment but certainly possible. It is especially risky for younger patients, who normally have a higher incidence of refractory symptoms. Studies conducted in recent years show proton treatment to be safer than photon therapy in this area \cite{neutrons}. The source of secondary cancer from retinoblastoma treatment was 1.8\% for proton therapy compared to 12.9\% for photon therapy, and 0\% compared to 7\% for medulloblastoma (brain cancer) \cite{neutrons}. Within the test group considered, all secondary cancer developed at sites away from the primary tumor, hence it is also possible that metastatic disease beyond the affected volume was initially missed.

The linear non-threshold theory (LNT) states that the probability of developing cancer is directly related to the amount of radiation exposure. This relationship has been established for doses $\geq$100 mSv but remains unknown for lower cases \cite{kino}. Both apoptosis (cellular self-destruction) and DNA repair give inhibition factors that change the probability of developing cancer. Some mathematical models suggest that for ranges up to 300 mSv (with threshold dependence greater than 750 mSv) certain cellular defense mechanisms might be inoculated by small amounts of radiation (hormesis). Low energy radiotherapy is generally a few Gy per treatment, and twenty-five or so are used in fractionation (NHS), so the threshold range is far above the levels speculated for hormesis. The tumor relapse rates, however, do remain low for the low energy test group.

\vspace{-3mm}
\section{Conclusion}
\vspace{-1mm}
Proton radiotherapy is becoming an increasingly popular alternative to conventional photon treatment, possibly serving some advantages to pediatric patients that have an increased likelihood of developing secondary cancer. In extreme metastatic cases, it is uncertain whether the benefit of excess radiation from photon beams outweighs the probability of developing radio-induced cancer. Since potential advantages of small radiation exposure are ill-understood, it remains advisable that radiation be administered to as little healthy tissue as possible lest secondary cancer be formed. For particularly localized cancers, especially those surrounded by critical anatomy, proton treatment is considered favorable to photon treatment due to its characteristic Bragg peak and immediate radiation fall-off. 

The current clinical challenges of proton therapy stem mostly from technical and financial issues associated with a lack of standardized imaging modalities, which are lacking in comparison to hospital linear accelerators. The Geant4 simulation showed that for reasonable proton energies, the energetic control over depth and low exit dose was more precise than for photons. For proton energies larger than 250 MeV, however, the radial profile of dose consumed more space than would be clinically acceptable, given a subject of the phantom-like parameters. For a bone-solid phantom, more energy was deposited into the bone material than into the water, showing that photon therapies in particular need to account for greater exit dose affecting those areas. 

\vspace{-2mm}
\printbibliography

@article{kino,
author = {Kino, Katsuhito},
year = {2020},
month = {02},
pages = {155-164},
title = {The Prospective Mathematical Idea Satisfying Both Radiation Hormesis Under Low Radiation Doses and Linear Non-threshold Theory Under High Radiation Doses},
volume = {42},
issue = {4},
journal = {Genes and Environment},
doi = {10.1186/s41021-020-0145-4}
}

@article{suit,
author = {Suit, Herman and Urie, Marcia},
year = {1992},
month = {02},
pages = {155-164},
title = {Proton Beams in Radiation Therapy},
volume = {84},
issue = {3},
journal = {Journal of the National Cancer Institute},
doi = {10.1093/jnci/84.3.155}
}

@article{cost,
author = {Verma, Vivek and Shah, Chirag  and Rwigema, Jean-Claude and Solberg, Timothy and Zhu, Xiaofeng and Simone II, Charles},
year = {2016},
month = {05},
title = {Cost-Comparativeness of Proton versus Photon Therapy},
volume = {5},
journal = {Chinese Clinical Oncology},
doi = {10.21037/cco.2016.06.03}
}

@article{groom,
author = {Groom, D.E. and Klein, S.R.},
year = {2019},
month = {08},
title = {Passage of Particles Through Matter},
journal = {The European Physical Journal C - Particles and Fields},
doi = {10.1007/BF02683419}
}

@article{kooy,
author = {Levin, W. and Kooy, H. and Loeffler, J.},
year = {2005},
month = {10},
title = {Proton Beam Therapy},
journal = {British Journal of Cancer},
doi = {10.1038/sj.bjc.6602754}
}

@article{gazis,
author = {Gazis, Evangelos},
year = {2019},
month = {01},
title = {The Ionizing Radiation Interaction with Matter, the X-ray Computed Tomography Imaging, the Nuclear Medicine SPECT, PET and PET-CT Tomography Imaging},
doi = {10.5772/intechopen.84356}
}

@misc{neutrons,
title = {Neutrons in Proton Therapy},
author = {Wei, Jonathan},
note = {Last accessed on: 11-03-2020}
}

\end{document}